\documentclass{css2022}
\usepackage{epsfig}
\usepackage{amssymb}	
\usepackage{hyperref}
\usepackage[none]{hyphenat}
\usepackage{xcolor}

\newcommand{\PACS}{\MSC}

\title{Local dynamical effects of scale invariance: the lunar recession }

\author{Andre Maeder$^1$, Vesselin Gueorguiev$^{2,3}$\\
\vskip 2mm {\small
$^1$
Geneva Observatory\\
chemin Pegasi 51, CH-1290 Sauverny, Switzerland\\
andre.maeder@unige.ch \\
$^2$ Institute for Advanced Physical Studies, Sofia, Bulgaria\\
$^3$ Ronin Institute for Independent Scholarship, Montclair, NJ, USA\\
Vesselin at MailAPS.org
}}

\abstract{Scale invariance  is  expected in empty Universe models, while the presence of matter tends to suppress it.
As shown recently, scale invariance is certainly absent in cosmological models with densities 
equal to or above the critical value $\varrho_{\mathrm{c}} =3H^2_0/(8 \pi G)$.
 For models with densities below $\varrho_{\mathrm{c}}$, 
the possibility of limited  effects remains open.
 If present, scale invariance would  be  a global cosmological
property.  Some traces could be observable locally. For the Earth-Moon two-body system, the predicted additional lunar recession 
would be increased by 0.92 cm/yr, while the tidal interaction would  also be slightly increased.\\
  The Earth-Moon distance is the most systematically  measured distance in the Solar System, thanks to
  the Lunar Laser Ranging  (LLR) experiment   active since 1970. 
 The observed lunar recession  from LLR  amounts to 3.83 ($\pm 0.009$) cm/yr;
 implying a  tidal change   of the length-of-the-day (LOD) by 2.395 ms/cy. However,  the observed change of the LOD 
 since the Babylonian Antiquity is only 1.78 ms/cy, a result supported by paleontological data, and
  implying  a lunar recession of  2.85 cm/yr. 
The significant difference  of  (3.83-2.85) cm/yr = 0.98 cm/yr,
already pointed out by several authors  over the last two decades, 
corresponds well to the predictions of the scale-invariant theory, 
which is also supported by  several other astrophysical  tests. 
}


\begin{document}

\maketitle

\section{Introduction} \label{intro}

 The scale-invariant theory aims at responding to a most fundamental principle expressed
by Dirac \cite{Dirac73}: {\emph{``It appears as one of the fundamental principles
in Nature that the equations expressing basic laws should be invariant
under the widest possible group of transformations''}}. Our objective is  to explore 
whether, in addition to Galilean invariance, Lorentz invariance, and  general covariance, some effects of
 scale invariance would also be present in our   low density Universe.
This is particularly justified since scale invariance is present in Maxwell's equations 
in absence of charges and currents, while in General Relativity (GR) 
 scale invariance is  a property of the empty space  in absence of a cosmological constant, a property  pointed out by \cite{Bondi90}. 
 
Clearly,  the presence of matter tends to kill  scale invariance as shown by \cite{Feynman63}.
Thus, the question arises about how much matter in the Universe is necessary for suppressing scale invariance.
Would one single atom in the Universe be enough to kill scale invariance?
 Clearly, we do not know whether this is the case, but the way to get an answer to the above questions is to
carefully examine the theoretical consequences of this assumption and to perform comparisons with
observations.  
From the scale covariant  expressions of the Ricci tensor, curvature scalar and general field equation
developped by   \cite{Dirac73} and \cite{Canuto77}, cosmological models were obtained by \cite{Maeder17a} 
who showed that scale invariance is clearly forbidden for   models 
with matter  densities equal and above the critical density
$\varrho_{\mathrm{c}}=3 H^2_0/(8 \pi G)$. This result was  found  consistent with  considerations 
\cite{MaedGueor21a} on causal connexion in Universe models.

        These models also indicate that, as soon that one considers models
with a density parameter $\Omega_{\mathrm{m}} > 0$,   scale-invariant effects are drastically reduced, before totally disappearing
at $\Omega_{\mathrm{m}} = 1$.  For a Universe model with 
$\Omega_{\mathrm{m}}=0.3$, they are nevertheless sufficient to drive a  significant acceleration of the expansion.
Several positive results  have  been obtained, \emph{e.g.} on  the  distance modulus vs. reshifts $z$ relation,
  the Hubble rate vs. age and density parameter, the $H(z)$ vs. $z$ relations, even if due to the observed scatter
   the discrimination from the $\Lambda$CDM is difficult at present
 \cite{Maeder17a}, \cite{MaedGueor20a}. The growth of density fluctuations is accounted for 
 without the need of dark matter \cite{MaedGueor19}; the same for 
 the observed mass excess in clusters of galaxies  \cite{Maeder17c};
 and  the radial acceleration relation for galaxies  is reproduced \cite{MaedGueor20b}.
For a brief summery of the Scale Invariant Vacuum paradigm and 
its main results and current progress see \cite{Gu+M'22}.

The question whether  astrophysical systems, such as the solar system and  galaxies, 
follow the Hubble-Lema\^{i}tre expansion has stimulated a vast literature since the pioneer work of 
McVittie \cite{McVittie32,McVittie33} and the Einstein-Straus theorem \cite{Einstein45}.  
The presence of an expansion at smaller scales
has been considered as an open question by Bonnor \cite{Bonnor00} and recently revisited by us \cite{MaedGueor21b}.
The fact that the dark-energy dominates the matter-energy content of the Universe and that this energy
appears as driving the acceleration of expansion is reviving the interest in the question:
If dark energy is uniformly distributed in space would it not imply effects that may be present at small scales?
The Earth-Moon system occupies a particular place in this context, since there are direct accurate measurements
of the evolution of the distance in this two-body system.
 
It is thus appropriate  to examine whether there are some  local effects of scale invariance, {\emph{e.g.}} in the Solar System. 
According to several pioneer works \cite{MBouvier79,Dumin03,Dumin16,Krizek16} the local effects in the Solar 
System due to scale invariance would  have been of the order of the Hubble-Lema\^{i}tre  expansion or some fraction of it.
(The Hubble-Lema\^{i}tre  expansion with $H_0=70$ km  s$^{-1}$ Mpc$^{-1}$
corresponds to 10.7 m/yr for one  astronomical unit, or 2.75 cm/yr for the Earth-Moon distance of 384'400 km).
For the proper interpretation of these effects in the scale-invariant context, it is necessary to
 account for the limitations of the $\lambda$-variations 
 due to a significant matter density in the Universe, and this implies  a refined analysis, which we do here.
 
Section \ref{recall} briefly recalls the main points of the scale-invariant vacuum idea. 
Section 3 examines the limitations of the scale-factor $\lambda$ and their impact on timescales. 
In Section 4, we study the weak-field low-velocity approximation of the scale-invariant field equation 
and the two-body problem. In Section 5,  we 
 compare the predicted and observed lunar recession from Lunar Laser Ranging (LLR) in relation with 
 the data on the length-of-the day (LOD).
Section 6 gives the conclusions.

\section{Some  points on the scale invariant vacuum (SIV) theory} \label{recall}

Some  recalls on the scale-invariant framework have been given recently \cite{MaedGueor21a}, with references therein. 
 In  short, the theory is expressed in the cotensor framework appropriate to the 
Integrable Weyl Geometry developed by \cite{Dirac73} and \cite{Canuto77}. The developpements are rather parallel to those of
General Relativity (GR), but with the possibility of  conformal scale transformations of the form,
\begin{equation}
ds'= \lambda(x^{\mu}) ds \,,
\label{ds}
\end{equation}
 in addition to the general covariance. Primed quantities refer to the GR framework, while quantities without a
 prime refer to the scale covariant context.
Scale covariant first and second derivatives, scale covariant Christoffel symbols, 
Riemann-Christoffel cotensor, Ricci cotensor and  total co-curvature have been  developed
 in the Integrable Weyl Geometry by \cite{Dirac73},  leading 
to a general scale covariant field equation \cite{Canuto77}.

In GR, one needs to define the line element corresponding to the physical system studied, for example the FLWR line element is adopted for 
expressing the cosmological equations in  a homogeneous and isotropic Universe.
Similarly, in the scale covariant context, an additional condition is necessary to fix the gauge. 
Dirac and Canuto et al. had chosen the then in vogue ``Large
Number Hypothesis`` \cite{Dirac74}. 
We prefer to adopt as  basic gauging  condition the following
 assumption: {\emph{The macroscopic empty space is scale invariant, homogeneous and isotropic}}.
 This is  a simple and most reasonable assumption, which is consistent with the scale invariance of the equations
 of  Maxwell  and of General Relativity in empty space, as  recalled in the introduction.
  Moreover, the  equation of state of the vacuum $p_{\mathrm{vac}}= -\varrho_{\mathrm{vac}}  c^2$  is precisely 
the one equation permitting  the vacuum density   to remain constant for an adiabatic
expansion or contraction  \cite{Carroll92}.  We also note that the assumption of homogeneity 
 and isotropy appears a  reasonnable one for the macroscopic empty space.

Within the cotensor framework, our gauging condition can be expressed as follows \cite{Maeder17a},
\begin{eqnarray}
  \kappa_{\mu ;\nu} + \kappa_{ \nu ;\mu}
+2 \kappa_{\mu} \kappa_ {\nu} - 2 g_{\mu \nu} \kappa^{ \alpha}_{;\alpha}
+ g_{\mu \nu}\kappa^{ \alpha} \kappa_{ \alpha} = 
  \Lambda \, g_{\mu \nu} \, .
\label{empty}
\end{eqnarray}
It is what is left from the scale covariant field equation if space is empty \footnote{
The de Sitter metric for empty space with $\Lambda_{\mathrm{E}}$ is conformal to the Minkowski metric,  
and is identical to it for the condition $3 \lambda^{-2}/(\Lambda_{\mathrm{E}} t^2)=c^2$. 
This condition is consistent with the solution (\ref{lambda}}.
The first member (LHS) results from the scale invariant form of the Ricci tensor \cite{Dirac73}.
The second member (RHS) contains only  the cosmological constant $\Lambda$ in the scale covariant form, with
\begin{equation}
\Lambda = \lambda^2 \Lambda_{\mathrm{E}}\, . 
  \end{equation}
  \noindent
 $\Lambda_{\mathrm{E}}$ is the cosmological constant in GR.
 The first member of Eq.(\ref{empty}) 
contains  terms depending on $\kappa_{\nu}$,  the coefficient of metrical connection, 
related to the scale factor $\lambda$ of Eq. (\ref{ds}),
\begin{equation}
\kappa_{\nu} \,  = \,
- \frac{\partial \ln \lambda}{\partial x^{\nu}} \, .
\label{ka}
\end{equation}     
\noindent
 We note that if the scale factor $\lambda$ is a constant, all terms in $\kappa_{\nu}$
  vanish and  Eq. (\ref{empty})  implies  $\Lambda_{\mathrm{E}}=0$. This  means 
  that the scale-invariant field equation  just becomes the field equation of GR without a cosmological constant.
  
For reasons of homogeneity and isotropy of the empty space,  the scale factor  $\lambda$  should depend on time only, 
so that the only non-zero component of $\kappa_{\nu}$ is $\kappa_0$,
\begin{equation}
\kappa_{\nu}=\kappa(t)\, \delta_{0 \nu},\;\kappa_{0, 0} = \frac{d \kappa_0}{dt} =\dot{\kappa}_0=\dot{\kappa} \,.
\label{k0}
\end{equation}
In Weyl's Integrable Geometry, $\kappa_{\nu}$ is playing a fundamental role alike the $g_{\mu \nu}$. 
From the time and space components of Equation (\ref{empty}) one obtains: 
\begin{eqnarray}
\  3 \, \frac{ \dot{\lambda}^2}{\lambda^2} \, =\, \lambda^2 \,\Lambda_{\mathrm{E}}  \,  
\quad \mathrm{and} \quad  \,  2\frac{\ddot{\lambda}}{\lambda} - \frac{ \dot{\lambda}^2}{\lambda^2} \, =
\, \lambda^2 \,\Lambda_{\mathrm{E}}\,. \label{diff1}
\end{eqnarray}
\noindent
Thus, the gauging conditions leads to  analytical relations between
the scale factor $\lambda$ and 
the cosmological constant, which represents the energy density of the vacuum.  
These differential equations 
give a new physical significance to the cosmological constant, which now  appears 
as  the energy density of the relative variations of the scale factor, see the first of Eqs.(\ref{diff1}) 
 and \cite{MaedGueor21a} for its connection to inflationary stage of the very early Universe.
The solution of these  differential equations is 
\begin{equation}
\lambda(t) \, =\, \sqrt{\frac{3}{\Lambda_{\mathrm{E}}}} \, \frac{1}{ct} \,.
\label{lambda}
\end{equation}
We take the present time $t_0=1$ and also consider the present scale as a reference to which all scales are referred to.
Thus, $\lambda(t)$ may be written as, 
\begin{equation}
{\lambda}(t)  \, = \, \frac{t_0}{t}\,.
\end{equation}

\begin{figure}[t]
\centering
\includegraphics[width=13.5cm, height=8.5cm]{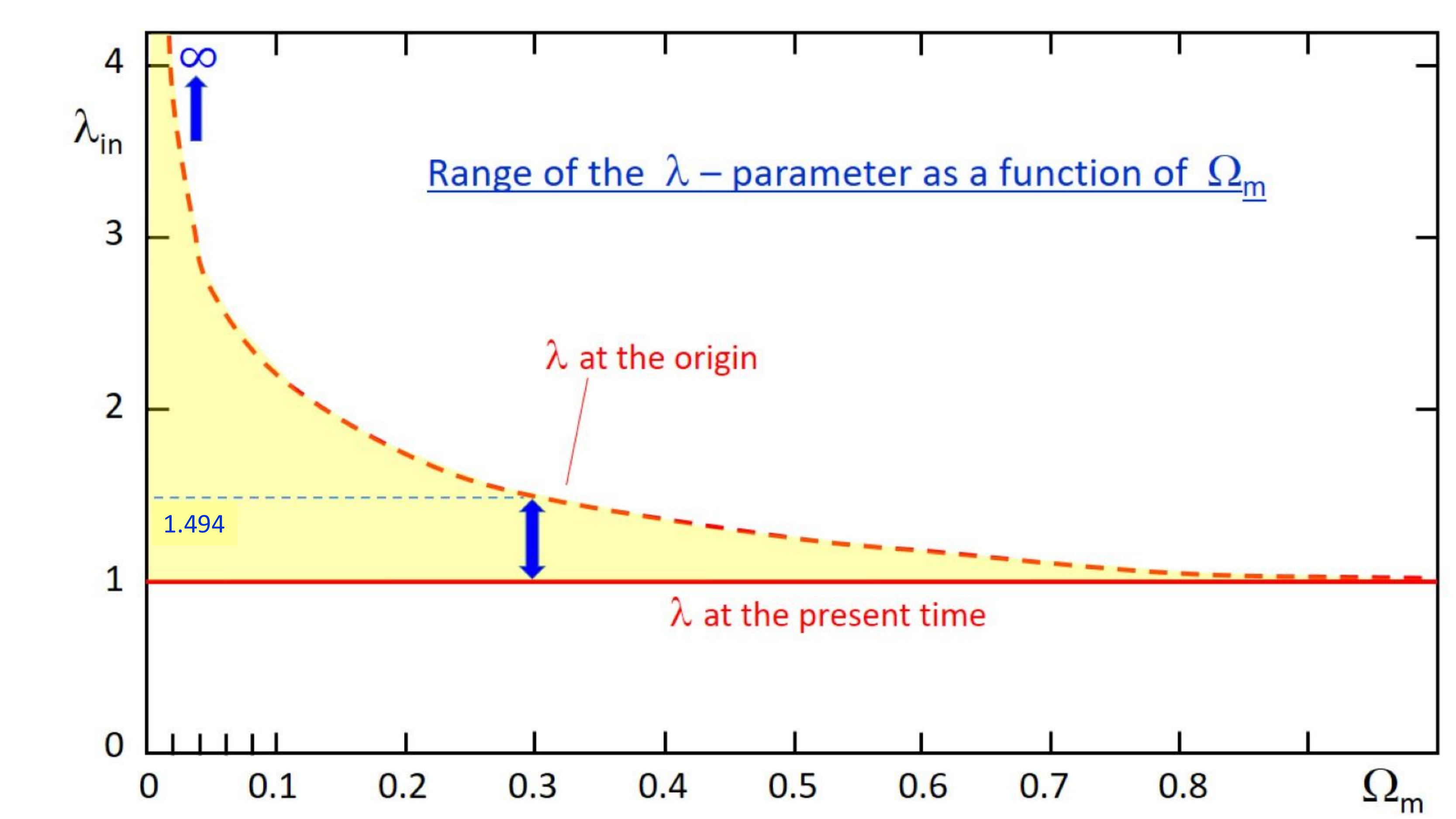}
\caption{The values of  the scale factor  $\lambda_{\mathrm{in}}= 1/t_{\mathrm{in}}$ at the initial time
 $t_{\mathrm{in}}=\Omega^{1/3}_{\mathrm{in}}$,  
as a function of the density parameter $\Omega_{\mathrm{m}}$. 
The yellow  zone shows, at each value of $\Omega_{\mathrm{m}}$,  the range of  $\lambda(t)$  
  from the Big-Bang (broken red line) to the present time (continuous red line).
The blue arrow illustrates  that for $\Omega_{\mathrm{m}}=0.3$, the value of $\lambda(t)$  varies only between 1.494 at the origin 
and 1.0 at present. We see the drastic  reduction of the effects of scale invariance 
with increasing $\Omega_{\mathrm{m}}$.
 }
\label{lambdin}
\end{figure}

Remarkably, 
the gauging condition, which implies the two equations (\ref{diff1}), 
lead  to major simplifications of the  cosmological equations derived by \cite{Canuto77} on the basis of the 
general field equations. One obtains \cite{Maeder17a}:
\begin{eqnarray}
\frac{8 \, \pi G \varrho }{3} = \frac{k}{a^2}+\frac{\dot{a}^2}{a^2}+ 2 \,\frac{\dot{a} \dot{\lambda}}{a \lambda} \, ,
\label{E1} \\
-8 \, \pi G p  = \frac{k}{a^2}+ 2 \frac{\ddot{a}}{a}+\frac{\dot{a^2}}{a^2}
+ 4 \frac{\dot{a} \dot{\lambda}}{a \lambda}  \, .
\label{E2}
\end{eqnarray}
\noindent
On the right side of both, we note an additional term. 
For a constant $\lambda$, Friedmann's equations are recovered.
 A  third equation may be derived from the above two,  
 \noindent
\begin{equation}
- \frac{4\pi G}{3} \left(3p+\varrho \right) = \frac{\ddot{a}}{a} + \frac{\dot{a} \dot{\lambda}}{a \lambda} \, .
\label{E3n}
\end{equation}
Since $\dot{\lambda}/{ \lambda}$ is negative,  the extra term represents  an additional acceleration in the direction of the 
motion. Thus, the effects of the scale invariance are fundamentally different from those of a cosmological constant.
For an expanding Universe, this extra force produces an accelerated expansion, without requiring 
 dark energy. For a contraction, 
 the additional  term  favors collapse. 
 This is exactly what the corresponding term in the weak-field approximation is also doing, as verified
  in the study of the growth of density fluctuations \cite{MaedGueor19}. There,  the
  additional term  favors contraction and allows 
 an early formation of galaxies in absence of dark matter.

 \section{Limits on the variations of the scale factor}
 
 \subsection{Solutions of the cosmological equations for $k=0$}     \label{kzero}
 
Analytical solutions  for the flat scale-invariant models with $k=0$  have been obtained by  \cite{Jesus18} 
in the case of matter dominated models, with the corresponding equation of conservation,
\begin{equation}
a(t) \, = \, \left[\frac{t^3 -\Omega_{\mathrm{m}}}{1 - \Omega_{\mathrm{m}}} \right]^{2/3}\, .
\label{R}
\end{equation}
It is noticeable that an analytical solution exists.
It is expressed in the timescale where at present  $t_0=1$ and  $a(t_0)=1$. Such solutions are illustrated in 
\cite{Maeder17a} and  \cite{MaedGueor21a}.
Along with $\Omega_{\mathrm{m}}= \varrho/\varrho_{\mathrm{c}}$  with $\varrho_{\mathrm{c}}=3H^2_0/(8\pi G)$
as usual.
There is no meaningful scale-invariant solution for $\Omega_{\mathrm{m}}$ equal or larger than 1, 
consistently with causality relations \cite{MaedGueor21a}.
We see that the initial time when $a=0$, the Big Bang, is at:
\begin{equation}
 t_{\mathrm{in}} \, = \, \Omega^{1/3}_{\mathrm{m}}\,.
 \label{t_in}
 \end{equation}
 This dependence in 1/3 produces a  rapid increase of $ t_{\mathrm{in}}$   for increasing  low matter density.
 For $\Omega_{\mathrm{m}}=0, 0.01, 0.1, 0.3, 0.5$, the values of  $ t_{\mathrm{in}}$ are 0, 0.215, 0.464, 0.669, 0.794.
 Since $\lambda \sim 1/t$, this leads to a strong reduction of the range of $\lambda(t)$-variations for increasing matter densities
 as illustrated in Fig. 1. While the range of scale variations is infinite between the Big Bang and now for an empty Universe model,
 the range would be very limited for significant $\Omega_{\mathrm{m}}$-values, for example from only 1.494 to 1.0 for 
 $\Omega_{\mathrm{m}}=0.3.$
 The Hubble parameter,  in the same timescale, is
 \begin{equation}
 H(t) = \frac{2 \, t^2}{t^3- \Omega_{\mathrm{m}}}\,.
 \label{H(t)}
 \end{equation}
 Thus, $H_0= H(t_0)$ varies between $2$ and the infinite for $\Omega_{\mathrm{m}}$ between 0 and 1.
We also write  
\begin{equation}
\Omega_{\mathrm{k}} = -\frac{k}{a^2  H_0^2} \, \quad \mathrm{and} 
\quad \Omega_{\lambda} \, =\,- \frac{2}{H_0} \left(\frac{\dot{\lambda}}{\lambda}\right)_0 \,= \,\frac{2}{ H_0 \, t_0} \,,
\label{hl}
\end{equation}
\noindent
These are respectively the normalized  contributions  (vs. $\varrho_{\mathrm{c}}$) 
of the matter, space curvature, and scale factor $\lambda$.
With these definitions equation (\ref{E1}) leads to, 
\begin {equation}
\Omega_{\mathrm{m}} \, + \, \Omega_{\mathrm{k}} \, +  \Omega_{\lambda} = \, 1  \, .
\label{Omegapr}
\end{equation}
\noindent
These quantities are usually considered
at the present time. 
In the case of energy-density dominated by radiation and relativistic matter, for flat scale-invariant models with $k=0$,
analytical solutions  for the expansion factor,
the matter density, the  radiation density and temperature have been obtained by \cite{Maeder19}.

\subsection{Some essential properties of the scale-invariant solutions}\label{3.2}

\noindent
Let us further examine the conditions of applicability of the above results:
\begin{enumerate}
\item[1.]  Case of $\lambda(t)$ for homogeneous and isotropic  Einstein-like empty space,
but  not necessarily empty Universe with homogeneous and isotropic cosmological space. 
The two equations (\ref{diff1}) and their solution (\ref{lambda}) for $\lambda(t)$ 
have been  derived for the macroscopic  empty space, 
under the assumption that it is homogeneous and isotropic, which implies a dependence of $\lambda$ on time $t$ only. 
The empty space obeys an equation $p_{\mathrm{vac}}= - \varrho_{\mathrm{vac}} \, c^2$ and in the scale-invariant 
theory  the vacuum density is also related to the cosmological constant by $\Lambda \, = \, 8 \pi G\, \varrho_{\mathrm{vac}}$.
For $\Omega_{\mathrm{m}}=0$ one has $\lambda=1/t$ and $a=t^2$ and 
the cosmological equation (\ref{E3n}) then implies $3p+\rho=0$,
which is the trace of the matter energy-momentum tensor. 

\end{enumerate}
\noindent
Alike in GR, the properties  of the vacuum and thus of the cosmological constant are 
intrinsic  characteristics of the vacuum space, not depending on any matter content
and  distribution, and so was it also in the derivation of Eqs. (\ref{diff1}).
Thus, these  equations and their solution  
apply everywhere in the Universe, independently of  $\Omega_{\mathrm{m}}$.
 As a consequence,  {\emph{ the solution $\lambda(t)$  is a universal function, characteristic of the empty space.}}
Let us note that  the energy density of the empty space  can  be expressed in term of a scalar field $\psi$,
 \begin{equation}
\varrho \, = \, \frac{1}{2}\, C \,  \dot{\psi}^2  \, \quad \mathrm{with}  
\; \,\dot{\psi}  \,= \,  \kappa_0 = \,- \frac{\dot{\lambda}}{\lambda} \, .
\label{ro2}
\end{equation}
with the constant $C= 3/(4\pi G)$. The field $\psi$ obeys a modified Klein-Gordon equation  and  
 $\psi$ is advantageously playing the role of  the ``inflaton'' during inflation \cite{MaedGueor21a}.   
\begin{enumerate}
\item[2.]  Case of $\lambda(t)$ in presence of matter. The presence of matter could be viewed as 
space inhomogeneities below certain scale  but absent at larger scales and especially at cosmic scales.
The presence of matter is  determined by the density parameter $\Omega_{\mathrm{m}} = 
 \varrho/\varrho_{\mathrm{c}}$, which
 influences the interval of time between  the initial time 
 $t_{\mathrm{in}}= \Omega^{1/3}_{\mathrm{m}}$ and $t_0=1$. 
   For a higher  $\Omega_{\mathrm{m}}$-value (between
 0 and  1), the interval $(t - t_{\mathrm{in}})$ is smaller and so does the range  of the $\lambda$-values. 
  In this indirect way,  the presence of matter  drastically  reduces  
   the range of variation of  the scale factor $\lambda$ (Fig. 1). 
Most importantly, we easily verify that in the scale-invariant context (Section \ref{kzero}), 
 $\Omega_{\mathrm{m}}$ does not change during the evolution of the Universe, 
 both in the matter and radiation eras.
 As long as the radiation era is very short and negligible compared to the full age of the Universe,
 then the initial time $t_\mathrm{in}$ defines $\Omega_{\mathrm{m}}$ as a constant for the Universe.
 \end{enumerate}
Thus, steps 1 and 2 show that the  function $\lambda(t)$ has a form which is universal  $\lambda(t)  \, \sim \, 1/t$,
  but the  range of the time variations, and thus of $\lambda$, is strongly limited by the matter content.
\begin{enumerate}
\item[3.]  Case of $\lambda(t)$ for comoving galaxies. 
Now, comouving galaxies  have  the same  timescale and thus the same age  (the cosmic time). 
Thus, the $\lambda(t)$ parameter, its first, second derivatives and the coefficient of metrical connection
are  the same in all comoving galaxies:
  \begin{equation}
  \kappa(t)\, =\,  - \dot{\lambda}/\lambda \, = \, 1/t\,.
  \end{equation}
\end{enumerate}
\noindent
Therefore, we are led to  the following  global conclusion:
 {\emph{The universal function $\lambda(t)$ and its limitations apply the same way in all comouving
 galaxies}}.
   Thus, the effects of the variations of $\lambda(t)$ with their  limitations  could also be expected locally, \emph{e.g.}
     in the Solar System, which is a low velocity  subpart of a comoving galaxy. 

\subsection{Relations between timescales}\label{timescales}
 
 We have two different timescales (both concerning the cosmic time): 
 (1.) The age $t$  of the above cosmological model,  
 with $t_0=1$ at the present time  and  $t_{\mathrm{in}} = \Omega^{1/3}_{\mathrm{m}}$ at the origin.
 (2.) The usual timescale $\tau$,  with  $\tau_0= 13.8$ Gyr at the present time \cite{Frie08} and  $\tau_{\mathrm{in }}=0$ at the Big-Bang.
 The relation between  ages in the two timescales  is,
\begin{equation}
\frac{\tau - \tau_{\mathrm{in}}}{\tau_0 - \tau_{\mathrm{in}}} = \frac{t - t_{\mathrm{in}}}{t_0 - t_{\mathrm{in}}}\, .
\end{equation}
This means that  for an event at a given epoch, the age fraction with respect to the present age is the same in both timescales.
This gives the two following relations between  particular times  $t$ and $\tau$,
\begin{equation}
\tau \,= \, \tau_0 \, \frac{t- t_\mathrm{in}}{t_0-t_\mathrm{in}} \, , \quad \mathrm{and} \; \;
  t \,= \, t_\mathrm{in} + \frac{\tau}{\tau_0} (t_0- t_\mathrm{in}) \,.
\label{T2}
\end{equation}
For the derivatives of these two timescales,  we have,
\begin{equation}
\frac{d\tau}{dt} \, = \, \frac{\tau_0}{t_0-t_\mathrm{in}}\,, \quad \mathrm{and}\; \;
\frac{dt}{d\tau} \, = \, \frac{t_0-t_\mathrm{in}}{\tau_0}\,.
\label{dT1}
\end{equation}
These derivatives have constant values, implying that the two times are linearly connected.
For larger  $\Omega_{\mathrm{m}}$-values, $t_\mathrm{in}$ is also larger (\ref{t_in}) and
the  timescale $t$ is squeezed over a smaller fraction of the interval [0,1] as  [$t_\mathrm{in}$, $t_0=1$]. 
The above expressions  are useful to express the  relative variations of the scale factor 
$\lambda(t)$ as function of ages.

\section{The dynamical equation and the two-body problem}

\subsection{The weak-field low-velocity approximation of the equation of motion}

 This approximation in the scale-invariant framework is leading to a modified expression of the Newton's Law
 \cite{MBouvier79,Maeder17c}. In spherical coordinates, it is 
  
 \begin{equation}
 \frac {d^2 \bf{r}}{dt^2} \, = \, - \frac{ G_t \, M(t)}{r^2} \, \frac{\bf{r}}{r}   +  \kappa(t) \,\frac{d\bf{r}}{dt} \, .
\label{Nvec}
\end{equation}
\noindent
This expression for a weak field is derived from the geodesic equation by Dirac
 in the reference \cite{Dirac73}, which was also obtained 
from an action principle by \cite{BouvierM78}.
The conservation law for a dust Universe in the scale 
invariant context imposes a relation of the form \cite{Maeder17a}:  $\varrho R^3 \lambda=const.$
This means that the  inertial mass of a particle or of an object  is not necessarily a constant, a situation which also occurs in GR and
Special Relativity where the inertial mass of an object depends on its velocity. 
Within SIV  the mass is evolving like $t$ in the same way  as the length, so that, interestingly enough, 
the gravitational potential  $\Phi =G_t \, M(t)/R(t)$ of an object is a scale-invariant quantity.
We call $M(t)$ the  mass that varies  like $M(t)  \sim  (1/\lambda) \sim t$,  where $t$ is  the cosmic time. 
In a Universe model with $\Omega_{\mathrm{m}} =0.3$, a mass $M(t_0)$ at the present time 
was  at time $t_{\mathrm{in}}= \Omega_{\mathrm{m}}^{1/3}$, 
$M(t_{\mathrm{in}}) = (t_{\mathrm{in}}/t_0) M(t_0)  = \Omega_{\mathrm{m}}^{1/3} \, M(t_0)= 0.6694 \; M(t_0)$ at the Big-Bang.
Over small and moderate  time intervals, the mass may often be considered as a constant.

 With respect to the classical expression, there is an additional acceleration term in the direction of motion. 
This term proportional to the velocity  is favoring collapse in an accretion system  and favoring expansion in a
two-body system.
From here onwards, we call $G_t$   the gravitational constant (a true constant), 
 expressed in the appropriate units  $t$ in the above cosmological models, 
 while  $G$ wil now on be reserved to the  value expressed in the current time units (years, seconds).

 Equation (\ref{Nvec})  contains  terms and derivatives which are   functions of the universal $\lambda(t)$, the range of which is squeezed 
 by the limited range of the adopted $t$-scale in the cosmological models. 
 As $\Omega_{\mathrm{m}}$  increases,  a given range $ \Delta \tau$ in the current time units (\emph{e.g.} 2 Gyr)  
  is expressed in terms of the corresponding smaller $\Delta t$  interval.
 We need to convert the  equation of motion (\ref{Nvec}) expressed with variable $t$ into terms of the cosmic time $\tau$
  in the current units (years, seconds) with an origin at $\tau=0$
 and a present value $\tau_0 =13.8$ Gyr. 
Equation (\ref{Nvec})  becomes,
 \begin{equation}
 \frac {d^2 \bf{r}}{d \tau^2} \left(\frac{d \tau}{dt}\right)^2 \, = \, - \frac{G_t \, M(t)}{r^2} \, \frac{\bf{r}}{r}   +
  \, \frac{1}{ t_\mathrm{in} +   \frac{\tau}{\tau_0} (t_0-t_\mathrm{in}) } \, \frac{d \tau}{dt}  \frac{d\bf{r}}{d\tau} \, \, .
\label{Nvec2}
\end{equation}
The corresponding units of $G$  in the usual $\tau$-scale are $ [cm^3\cdot g^{-1} s^{-2}]$. Thus, we have the correspondence
$G_t \,  \left(\frac{dt}{d\tau}\right)^2 = G$ with the usual units conversion. At the present epoch  $t_0$ or $\tau_0$ in the current units,
the masses $M(t_0)$ and  $M(\tau_0)$ are evidently equal. At other epochs, the relation is,
\begin{equation}
M(t)\, = \, \frac{t}{t_0} \, M(t_0) \, ,  \quad  \quad \mathrm{thus} \quad 
M(\tau)\,=\, [ \Omega^{1/3}_{\mathrm{m}} + \frac{\tau}{\tau_0} (1- \Omega^{1/3}_{\mathrm{m}})] \, M(\tau_0) \, .
\end{equation}
This results in the correct scaling of expected mass at the Big Bang $M(t_\mathrm{in})\, =\Omega_\mathrm{m}^{1/3}M(t_{0})$ 
to be compared with $M(\tau_\mathrm{in}=0)\, =\Omega_\mathrm{m}^{1/3}M(\tau_{0})$; therefore, 
$M(t_\mathrm{in})\, =M(\tau_\mathrm{in}=0)$.

Thus, multiplying both members of Eq.(\ref{Nvec2}) by $\left(\frac{dt}{d \tau}\right)^2$, we get at time  $\tau/\tau_0$,
 \begin{equation}
 \frac {d^2 \bf{r}}{d \tau^2}  \, = \, - \frac{G \, M(\tau)}{r^2} \, \frac{\bf{r}}{r}   
+ \, \frac{1}{ t_\mathrm{in} +   \frac{\tau}{\tau_0} (t_0-t_\mathrm{in}) } \, \frac{t_0-t_\mathrm{in}}{\tau_0}\,\frac{d\bf{r}}{d\tau} \, \, .
\label{Nvec3}
\end{equation}
We see that the presence of matter ($t_\mathrm{in}>0$) always reduces the effect of the additional acceleration term, 
for $\Omega_{\mathrm{m}}=1 \Rightarrow t_\mathrm{in}=1=t_0$ it disappears.
Now, let us consider the situation at the present time $\tau_0$. We define the numerical factor $\psi=\psi(\tau_0)$ using:
\begin{equation}
\psi(\tau) \, = \, 
\frac{t_0-t_\mathrm{in}}{ t_\mathrm{in} +   \frac{\tau}{\tau_0} (t_0-t_\mathrm{in}) }\;
\Rightarrow
\psi_0=\psi(\tau_0) \, =1-\Omega^{1/3}_{\mathrm{m}} \,.
\label{phi}
\end{equation} 
The modified Newton's equation at present time $\tau_0$, currently 13.8 Gyr, is then:
\begin{equation}
 \frac {d^2 \bf{r}}{d \tau^2}  \, = \, - \frac{G \, M(\tau_0)}{r^2} \, \frac{\bf{r}}{r}   + \frac{\psi_0}{\tau_0}   \frac{d\bf{r}}{d\tau} \, \,.  
\label{Nvec4}
\end{equation}
The small additional term depends on the global cosmology, in particular on parameter  $\Omega_{\mathrm{m}}$,
 as discussed above.
In an empty Universe, $\psi_0= 1$. For  $\Omega_{\mathrm{m}} \rightarrow 1$, one has $\psi_0 \rightarrow 0$.
 For higher $\Omega_{\mathrm{m}}$ values,
consistently with  the discussion of the  
Eq. (\ref{R}) the additional term would be absent \cite{MaedGueor21a}.
In  the case with $\Omega_{\mathrm{m}} = 0.30$, one has $\psi_0=0.331$. Thus, the additional acceleration term is significantly reduced.

\subsection{The two-body problem with $\lambda$-limitations}
Let us now consider   a two-body system within the SIV theory 
with $k=0$  and a density parameter $\Omega_{\mathrm{m}}$. From Eq.(\ref{Nvec}) in the $t$-scale,
  the orbital motion was found to be still described by the Binet equation (the mass change being accounted for)  and thus obeying 
  the equation of conics \cite{MBouvier79} and \cite{Maeder17c}
  \begin{equation}
  r(\vartheta) \, = \, \frac{p}{1+e\; cos \vartheta}\,.
  \end{equation}
  The parameter $p$ is related to the semi-major axis $a$, semi-minor axis $b$, and eccentricity $e$ via the relationships:
  \begin{equation}
  p \, =\frac{b^2}{a}\,,  \quad   b\, = \, a \sqrt{1-e^2}\,, \quad \mathrm{thus} \; \; p\,=\, a(1-e^2)\,.
  \end{equation}
  The eccentricity $e$  is a scale-invariant quantity. For $e=0$, one has a circular orbit with a  radius $r =p$.
There is a small  growth  of the parameter $p$, or  $r$ for $e=0$ as we consider here, first in the time $t$-scale,
\begin{equation} 
\frac{\dot{r}}{r} \, = (-\frac{\dot{\lambda}}{\lambda} ) \,= \, 1/t \, \quad \mathrm{implying} \;
r \, \sim \, t \, ,  \; \mathrm{and} \; \frac{\Delta r}{r} \, = \, \frac{\Delta t}{t} \,.
\label{tr}
\end{equation}
Thus, the  orbital motion  of a bound system is described by a circle (or an ellipse)
with a  slight superposed outwards spiraling  motion. 
Quite interestingly, the small cosmological expansion keeps the orbital velocity constant.
This reminiscent of the behaviour in MOND,
where the speed on an orbit becomes asymptotically
independent of the size of the orbit  \cite{Milgrom14}.

The above expressions formally only apply within an empty Universe with  $\Omega_{\mathrm{m}}=0.$
The limitations of the range of $t$- and $\lambda$-variations 
given by Eqs. (\ref{T2}) are not yet included. 
Clearly, we have to account for them
in a Universe with a density parameter different from zero. 
Thus, with  Eqs. (\ref{T2}) and (\ref{dT1}) we may write the relative change of the orbital radius (or parameter $p$),
\begin{eqnarray} 
\frac{dr}{r} \, = \, \frac{dt}{t} \, = \, \frac{dt}{d\tau} \, \frac{d\tau}{t} \,=\, \frac{(t_0-t_{\mathrm{in}})}{\tau_0}
\frac{d\tau}{(t_{\mathrm{in}} + \frac{\tau}{\tau_0} (t_0- t_{\mathrm{in}}))} =   \psi(\tau)\frac{d\tau\,}{\tau_0} .
\end{eqnarray}
\noindent
This applies at a time $\tau$. 
Let us consider, at the present epoch $\tau_0$ where the orbital radius is $r_0$, an interval of time $\Delta \tau$  (say one year) 
very small with respect to $\tau_0 =13.8$ Gyr.
We may thus write the relative change of the orbital distance  $\Delta r/\Delta \tau$ during this small interval of time,
\begin{equation}
\frac{\Delta r}{\Delta \tau} \, = \, \psi_0 \, \frac{ r_0}{\tau_0} \,.
\label{Deltar}
\end{equation}
For $\Omega_{\mathrm{m}}=0$,  we would get $\psi_0=1$ and thus $\Delta r/\Delta \tau = r_0/\tau_0$.
In a Universe model with a density parameter $\Omega_{\mathrm{m}} > 0$,
the  temporal increase   of the orbital parameter is smaller than  that in an empty universe 
($\psi_0=1-\Omega_{\mathrm{m}}^{1/3}<1$).
For $\Omega_{\mathrm{m}}$ tending to 1,  the relative orbital increase  tends to zero.
For  $\Omega_{\mathrm{m}}=0.2, 0.3, 0.4$, the factor $\psi_0$ is equal to 0.4152, 0.3306, 0.2632
respectively. Below, we will consider the  standard model with $\Omega_{\mathrm{m}}=0.30$.
On the whole, the account of  matter  strongly reduces the expected effects of scale invariance.

\section{Study of the Earth-Moon system}

 \subsection{The LLR data}
   Let us turn to the observations. 
Since  March 1970, the Earth-Moon distance has been intensively measured by
Lunar Laser Ranging (LLR), first at Mc Donald Observatory and since the 80's at several other observatories
around the world. A total of 20'138 ranges up to September 2015 have been performed leading to
 an average lunar recession of 3.83 $ (\pm 0.009)$ cm/yr \cite{Williams16a,Williams16b}. We note the impressive accuracy.
  The value of the lunar recession has not much changed since the first determination  
more than  three decades ago \cite{Christodoulidis88}, which illustrates the  quality of the measurements.
The Earth-Moon distance is the most intensively and systematically measured distance  in the Solar System 
and the only one  by a direct laser signal. 

 The main effect producing  this recession is the  Earth-Moon tidal coupling: 
 the tidal bump of the Earth embarked by the  fast Earth axial  rotation, with an angular velocity  faster than that of  the Moon
 on its orbits,  generates a forward pull on the Moon. This pull is transferring some axial terrestrial angular momentum to the lunar orbital one.
 Over geological times, the lunar recession has likely  changed since the Moon was closer to the Earth and thus the tidal
 exchanges were larger. For example, a lunar recession of about 6.5 cm/yr was estimated for the  Ediacaran - Early Cambrian 
 period (about 600 Myr ago) by \cite{Azarevich17}.
 
 \subsection{The LOD data} 
 
 According to the theoretical modeling of the tidal waves, lunar motion, and terrestrial rotation  by  \cite{Williams16a},
  the observed lunar recession of 3.83 cm/yr implies 
 an increase of the length of the day (LOD) of 2.395 ms/cy (millisecond per century). 
 The LOD is defined as the excess of the length of the day with
 respect to 86'400 SI seconds. We note that the   value of 2.395 ms/yr for the change of the LOD over centuries
 is  consistent with the calculation of the tidal coupling under the assumption of 
 conservation of the global angular momentum of the Earth-Moon system \cite{Dumin03,MaedGueor21b}. 
  This theoretical value of  the increase of the LOD has also been  very well confirmed recently by
 \cite{Baenas21}. These authors revisited the Earth rotation theories
 with a two-layer deformable Earth model, including dissipative effects
 at the core-mantle boundary and account of the coupling torque between the two.
  The deceleration is numerically estimated with frequency-dependent modeling of the various solid and oceanic
 tides. In the assumption of the long-term coupling of the core
 and mantle, they obtain a deceleration of the Earth rotation
 corresponding to an increase of the LOD of 2.418 ms/cy in very good agreement with the above modeling by \cite{Williams16a}.

 The best and longest studies on the change of the LOD in History have been performed by Stephenson et al.  \cite{Stephenson16},
 who  analyzed the lunar and solar eclipses from 720 BC up to 1600 AD and found an average  shift of the LOD by 1.78 ($\pm$ 0.03) ms/cy.  
 Such a shift,  apparently very small, is acting every day and progressively produces  large effects  over a long period.
 When cumulated over 2000 yr, differences in the time of solar and lunar  eclipses are  amounting  to  about 18'000 s. 
 Such  time differences  are leading  to big shifts in the eclipse locations, up to thousands of km.  
 The constraints on the eclipse locations and  indications  (when available) about the time of the eclipse allowed Stephenson et al. to make
 the above determination, which had only slightly changed  in  their successive papers since  1984, see references in \cite{Stephenson16}.
These authors also  suggested the existence of a slight undulation around the mean with 
  a period of 1500 yr, possibly related to the effect of magnetic core-envelope coupling  \cite{Dumberry06}.
  
   Various other effects may contribute to the LOD, with different timescales: atmospheric effects,
 ice melting and change of the sea level, glacial isostatic adjustment, and core-envelope coupling. 
 The most uncertain contribution is that of the core 
 coupling which seems responsible for the 1500 yr undulation.
 Over the long term, the various negative  and positive effects appear to  balance each other 
 before the year 1990, see data by  \cite{Mitrovica11,Mitrovica15} and our recent  detailed review of the problem  \cite{MaedGueor21b}, 
  (since 1990, the fast melting of even the polar ice fields 
  contribute to an increase of the Earth momentum of inertia).
 The reality of the difference 
 between the above observed mean value of the LOD (1.78 ms/cy) and the value due to the tidal interaction
 (2.395 ms/cy) has recently been further emphasized by \cite{Stephenson20}.
 
 We note that other estimates of the change of the LOD have been made from  lunar occultations,
 however on a shorter time basis. The observations of lunar occultations from 1656 to 1986 
 have been analyzed by \cite{McCarthy86}, they indicate a slowing down of the Earth of
 0.73 ($\pm 0.018$)  ms/cy. The data show some  decadal
 variations around the mean, which were also present in the further analyses. Various astronomical telescopic
observations over the last 350 yr were analyzed by \cite{Sidorenkov97}, which led to a mean increase of the LOD of
0.9 ms/cy. A new determination of the change of the LOD based on the data from \cite{Stephenson16} for
 occultations since 1680 complemented by IERS data for the period 1970 to 2020
has  been performed recently by us  \cite{MaedGueor21b}. We showed 
an average increase  of 1.09  ($\pm 0.012$) ms/cy for the LOD over
that period. All these determinations based on relatively  limited durations are more subject to decadal fluctuations
than the ancient eclipse observations. 
 
 Interestingly enough, another  independent very long-term determination of the deceleration of the Earth rotation  
 has been established on the basis of paleontological
 studies by Deines and Williams \cite{Deines16}. These authors examined all available paleontological fossils  and deposits for direct measurements of
 Earth's rotation, in particular they used corals, bivalves, brachipods, rythmites and stromatolites. 
 These fossils 
 are keeping  the  traces of phases of daily growth due to the alternance of days and nights.
 The oldest records go back to 1.85 Gyr ago, however such very ancient data are highly uncertain.  Much more reliable data are availablle since 
  the Cambrian explosion of the forms of life, when animals with hard shells first appeared, 
  about 542 Myr ago. From their whole sample, \cite{Deines16} found
  a clear decrease of the number of days in one year.  For example, 
  400 Myr ago, an epoch  where there are  lots of data, 
  the mean number of days in one year  (considered of constant duration, but see below) 
  was about $405\;$days (with an uncertainty $\sigma$
  of about $\pm$ 7 days). 
   From their sample of collected data, Deines and Williams found a mean deceleration
  corresponding  to a change of the LOD 1.642 ($\pm$ 0.48)  ms/cy.
This mean value is quite interesting, although not taken at present time 
it concerns an age differing by less 3\% of the present age of the Universe. 
The error is larger than the one by Stephenson et al. \cite{Stephenson16} ($1.78\pm0.03$ ms/cy). 
Nevertheless, it gives another value  consistent with and independent  of the value by Stephenson et al. \cite{Stephenson16}. 
Based  on an incredibly much longer time period, the result by Deines and Williams is supporting the discrepancy between
the change of the LOD and the value of the lunar recession.
  
\subsection{Theoretical Predictions  of the two-body problem}

Let us  consider the possible increase of the Earth-Moon distance during one year in the scale-invariant context.  
For an empty Universe  with an age   of $\tau_0=13.8$ Gyr and a mean Earth-Moon  distance of $r_0=384 400$ km,
the lunar recession  would  be
\begin{equation}
\left(\frac{\Delta r}{\Delta \tau}\right)_{\mathrm{vac}} \, = \, \frac{r_0}{\tau_0} \, = \,2.78  \;  \mathrm{cm \cdot yr}^{-1} \,.
\label{Deltaz}
\end{equation}
This is close to the Hubble-Lema\^{i}tre expansion rate 
with  $H_0=70$ km/(s Mpc), corresponding to a lunar recession of 2.75 $\mathrm{cm \cdot yr}^{-1}$.
For a Universe with   $\Omega_{\mathrm{m}}=0.30$, $\psi_0=0.3306$
 and the  predicted cosmological expansion of the Earth-Moon system is,
\begin{equation}
\left(\frac{\Delta r}{\Delta \tau}\right)_{\mathrm{cosm}} \, = \, 0.92 \; \mathrm{cm \cdot yr}^{-1} \,,
\label{prediction}
\end{equation}
a value substantially smaller than  the Hubble-Lema\^{i}tre  expansion. 

Let us also  shortly consider the Earth-Mars distance.
    The relevant parameters  are:
    
    \noindent
  Mars : $a=227 944 000$ km, $e=0.09339$, $p=225.956 \cdot 10^6$ km.\\
  \noindent
  Earth: $ a= 149 598 000$ km, $e= 0.01671$, $p= 149.556 \cdot 10^6$ km.\\
  
  \noindent
  The difference of parameters $p$
  for Mars and  the Earth  is $ d=p_{\mathrm{Mars}} - p_{\mathrm{Earth}}= 76.4 \cdot  10^6$ km.
  The estimate of  the Mars-Earth recession based on this distance $d$   would be in  an empty Universe, 
 $\left(\frac{\Delta d}{\Delta \tau}\right)_{\mathrm{vac}}\, = \,  \frac{ d}{\tau_0} \, = \,5.54  \;  \mathrm{m \cdot yr}^{-1}.$
With the factor $\psi=0.3306$ for $\Omega_{\mathrm{m}}=0.30$, we get
\begin{equation}
\left(\frac{\Delta d}{\Delta \tau}\right)_{\mathrm{cosm}} = \, 1.83 \, \mathrm{m \cdot yr}^{-1} \,.
\end{equation}


\subsection{The tidal interaction in the scale invariant context}

Let us also examine the tidal interaction in the Earth-Moon system.
The law of angular momentum conservation  for a  given mass element in the scale invariant framework \cite{MBouvier79},
is $\kappa(t) \, r^2  \, \Omega = const$,
while both $r$ and $M$, are scaling like $t$, e.g. $M=M_0 (t/t_0)$.
Let us  examine the conservation of the total angular  momentum of the Earth (E)--Moon (M) system 
at time $t$ \cite{Dumin03},
\begin{equation}
\zeta \,\cos \varphi \, I_{\mathrm{E}} \Omega_{\mathrm{E}}
+ M_{\mathrm{M}}  R^2 \Omega_{\mathrm{M}} = L  \,,  \quad  \mathrm{with} \; L=L_0 \, \frac{t^2}{t_0^2} \,      
\label{cons}
\end{equation}
The angle $\varphi$
is the variable angle between the lunar orbital plane  and  the Earth equator. 
 The numerical factor $\zeta$ accounts for the 
consequences of the eccentricity $e=0.055$ of the lunar orbit, see numerical value below.
Quantities $I_{\mathrm{E}}$ and $\Omega_{\mathrm{E}}= 2 \pi /T_{\mathrm{E}} $ are  
respectively the moment of inertia and the axial angular velocity of the Earth.
$M_{\mathrm{M}} $ is the mass of the Moon and $\Omega_{\mathrm{M}}$ its
orbital angular velocity. $R$ is the mean distance between the Earth and Moon. 
$L_0$ is the total angular momentum at the present time $t_0$. 
We neglect the axial angular momentum of the Moon, 
since its mass is 1.2 \% of that of the Earth and its axial rotation period (equal to its orbital period) is 27.3 days. 
Thus, the lunar axial angular momentum is a fraction of about $4 \cdot 10^{-4}$ of that of the Earth.

Let us evaluate the time derivative of the above expression (\ref{cons}),
\begin{eqnarray}
-\zeta \, \cos \varphi \,\frac{2 \pi}{T^2_{\mathrm{E}}}  I_{\mathrm{E}} \frac{dT_{\mathrm{E}}}{dt} +
\zeta \,\cos \varphi  \, \frac{2 \pi}{T_{\mathrm{E}}}  \frac{d{I}_{\mathrm{E}}}{dt} +
 \frac{d}{dt} (M_{\mathrm{M}} R^2 \Omega_{\mathrm{M}} ) =  \nonumber \\
2 \, \left(\zeta \, \cos \varphi \, I_{\mathrm{E}} \Omega_{\mathrm{E}}+
M_{\mathrm{M}}R^2 \Omega_{\mathrm{M}}  \right)_0 \frac{t}{t^2_0} \,.
\label{der1}
\end{eqnarray}
Account  has been given to the  change of  the moment of inertia due to the mass variation.
 $T_{\mathrm{E}}$ is the axial rotation period of the  Earth.
For $\Omega_{\mathrm{M}}$,  we have the relation
$ \Omega_{\mathrm{M}}^2=  G M_{\mathrm{E}}R^{-3}$,
 which also applies in the SIV context, there $M_{\mathrm{E}}$ is the Earth's mass. 
We can develop the third term on the left of the above expression (\ref{der1}),
\begin{eqnarray}
 \frac{d}{dt} (M_{\mathrm{M}} R^2 \Omega_{\mathrm{M}} ) = \frac{d}{dt}(G^{\frac{1}{2}}
M^{\frac{1}{2}}_{\mathrm{E}} R^{\frac{1}{2}} M_{\mathrm{M}} ) = 
G^{\frac{1}{2}}  M^{\frac{1}{2}}_{\mathrm{E}} M_{\mathrm{M}}\frac{d}{dt}(R^{\frac{1}{2}})+
G^{\frac{1}{2}}  R^{\frac{1}{2}}\frac{d}{dt}(M^{\frac{1}{2}}_{\mathrm{E}} M_{\mathrm{M}} ) = \nonumber\\
G^{\frac{1}{2}}  M^{\frac{1}{2}}_{\mathrm{E}} M_{\mathrm{M}}R^{-\frac{1}{2}}\frac{1}{2}\frac{dR}{dt}+
\frac{3}{2} G^{\frac{1}{2}} M^{\frac{1}{2}}_{\mathrm{E}} M_{\mathrm{M}} R^{\frac{1}{2}} \, \frac{1}{t}\, .\nonumber\\
\label{terme3}
\end{eqnarray}
Indeed, $dM/dt= M/t$ as well as $ M_0/t_0$.
Introducing this expression in  Eq.(\ref{der1}) and explicating  the time dependence of the various terms  leads to
\begin{eqnarray}
\zeta \,\cos \varphi  \, \left(\frac{2\,\pi}{T_{\mathrm{E}}} \, \frac{dI_{\mathrm{E}}}{dt}
- \frac{2 \,\pi  I_{\mathrm{E}}}{T^2_{\mathrm{E}}} \, \frac{dT_{\mathrm{E}}}{dt} \right) + 
\frac{G^{\frac{1}{2}} M^{\frac{1}{2}}_{\mathrm{E}} \, M_{\mathrm{M}}}{2 R^{\frac{1}{2}} }\, \frac{dR}{dt}+   
\frac{3}{2} G^{\frac{1}{2}} M^{\frac{1}{2}}_{\mathrm{E}} \, M_{\mathrm{M}} \,R^{\frac{1}{2}} \, \frac{1}{t}\, = \nonumber \\
\zeta \,\cos \varphi \, I_{\mathrm{E}} \,\frac{4 \, \pi}{T_{\mathrm{E}}} \, \frac{t}{t^2_0}+
2  G^{\frac{1}{2}} M^{\frac{1}{2}}_{\mathrm{E}} \,M_{\mathrm{M}} \,R^{\frac{1}{2}} \,\frac{t}{t^2_0} \, .
\label{der2}
\end{eqnarray} 
From this relation, we now extract  the lunar recession $dR/dt$, which  becomes after some simplifications, 
\begin{eqnarray}
\frac{dR}{dt}=\frac{4 \,\pi \zeta \, \cos\varphi \, R^{\frac{1}{2}} \, I_{\mathrm{E}}}
{G^{\frac{1}{2}} \,M^{\frac{1}{2}}_{\mathrm{E}}\,M_{\mathrm{M}} \,T^2_{\mathrm{E}}}\, \frac{dT_{\mathrm{E}}}{dt} + 
\frac{4\, \pi \zeta \, \cos\varphi  R^{\frac{1}{2}}I_{\mathrm{E}}}
{G^{\frac{1}{2}} M^{\frac{1}{2}}_{\mathrm{E}} \,M_{\mathrm{M}} \, T_{\mathrm{E}}}\,
\left(\frac{2t}{t_0^2}- \frac{1}{I_\mathrm{E}}\frac{dI_\mathrm{E}}{dt} \right)+ R\left(\frac{4t}{t_0^2} -\frac{3}{t}\right)\, .
\label{dr}
\end{eqnarray}
We may write it in a more condensed form by defining 
 a constant $k_{\mathrm{E}}$,
\begin{equation}
k_{\mathrm{E}} \, = \,4 \,\pi \left(\zeta \,\cos \varphi \frac{R^{1/2} \,I_{\mathrm{E}}}
{T^2_{\mathrm{E}} \,G^{1/2}\,M_{\mathrm{M}}\,M^{1/2}_{\mathrm{E}}}\right).
\end{equation} 
By using $dI_{\mathrm{E}}/dt=  3 \, I_{\mathrm{E}}/t$ (since $I_{\mathrm{E}}$ is scaling 
like $I_{\mathrm{E}}= I_0  (t^3/t^3_0)$  equation (\ref{dr}) can now be written as:
\begin{equation}
\frac{dR}{dt} \, = \, k _{\mathrm{E}} \, \frac{dT_{\mathrm{E}}}{dt}+ 
 k _{\mathrm{E}} T_{\mathrm{E}} \left(\frac{2 \,t}{t_0^2}- \frac{3 \,t^2} {t_0^3}\right)+ 
R\left(\frac{4t}{t_0^2} -\frac{3}{t}\right)\,\,.
\end{equation}
For  a time $t$  differing  very little from $t_0$, {\emph{e.g.}} by one year, one has  
$ \frac{t_0}{t} \rightarrow 1$, and thus with a high accuracy we write,

\begin{equation}
\frac{dR}{dt} \, =  \, k_{\mathrm{E}} \, \frac{dT_{\mathrm{E}}}{dt} - 
 k _{\mathrm{E}} \, \frac{T_{\mathrm{E}}}{t_0} +\frac{R}{t_0}\, .
\label{fin1}
\end{equation}

The first term on the right represents the tidal effect  linking the change of the LOD and the lunar recession, the second term
results from the increase of   the moment of inertia of the Earth  (which reduces the lunar recession), the third term expresses the global expansion of the 
system. Interestingly enough,  this third term is just the same as that predicted by the study of  to  the two-body 
problem, which shows the internal consistency of both approaches. We note that the  change of the mass of the Moon 
is also contained in this third term, it was intervening 
through  the last term in Eq.(\ref{dr}).

We have now to bring the above equation in the current time units, seconds and years. 
With Eq. (\ref{T2}) and (\ref{dT1}), we get
\begin{equation}
\frac{dR}{d\tau} \, = \, k_{\mathrm{E}} \, \frac{dT_{\mathrm{E}}}{d\tau}- 
 k _{\mathrm{E}} \,   \psi _0\,\frac{T_{\mathrm{E}}}{\tau_0} +
 \psi_0  \,\frac{R}{\tau_0}\, .
\label{fin}
\end{equation}
In a cosmological Universe model with $\Omega_{\mathrm{m}}=0.30$, the ratio $\psi_0 = \frac{(t_0-t_{\mathrm{in}})}{t_0} \,=0.331$.
The constant $k_{\mathrm{E}}$  is here in the units of $t^{-1}$, while  $T_{\mathrm{E}}$ is in the units of $t$.
Thus, both have to be turned $\tau$-units and the scaling factors simplify. Thus, $k_{\mathrm{E}}$  and $T_{\mathrm{E}}$
can finally be both  expressed in current  time units (seconds or years).

\subsection{Numerical values and discussion}

We adopt the following numerical values of the various astronomical quantities
\begin{eqnarray}
M_{\mathrm{E}} = 5.973 \cdot 10^{27} \mathrm{g},  \quad  \quad R_{\mathrm{E}}= 6.371 \cdot 10^{8} \mathrm{cm}, \\ \nonumber
\; \; \quad M_{\mathrm{M}}= 7.342 \cdot 10^{25} \mathrm {g},\quad \quad R=  3.844 \cdot 10^{10} \mathrm{cm},\\ \nonumber
\quad I_{\mathrm{E}} = 0.331 \cdot  M_{\mathrm{E}}   R^2_{\mathrm{E}} =
8.0184 \cdot  10^{44} \mathrm{g \cdot cm}^2. \nonumber
\end{eqnarray}
The value 0.331  is obtained from precession data \cite{Williams94}.
Coefficient $k_{\mathrm{E}}= 1.806 \cdot 10^5 \cdot \zeta \cos \varphi\; {\rm cm} \cdot {\rm s}^{-1 }$. 
The angle $\varphi$ varies between 18.16  and 28.72 degrees, thus we adopt a mean value of $\cos \varphi = 0.91$.
We have to estimate the value of $\zeta$. 
The tidal effects behave like $1/r^3$, thus they depend  on the eccentricity like 
$(1+e \, \cos \vartheta)^3$. 
The time spent in an interval of angles $\Delta \vartheta$ goes like $r^2 \Delta  \vartheta$, 
thus like $(1+e \, \cos \vartheta)^{-2}$.  
The product of the two  leads to a dependence of the form $(1+e \, \cos \vartheta)$.
For $\cos \vartheta$, let us take a value of $-0.5$ which leads to $\zeta=1 - 0.03 \approx 0.97$. We thus obtain for the 
 coefficient $k_{\mathrm{E}}= 1.60 \cdot 10^5$ cm $\cdot$ s$^{-1 }$.

Let us evaluate numerically the various   contributions. 
With the LOD  of 1.78 ms/cy from the antique data by \cite{Stephenson16}, the first term contributes to a lunar recession of 2.85 cm/yr,
while the LOD of 1.09 ms/cy  from 1680 to the present \cite{MaedGueor21b} leads to a recession of 1.74 cm/yr.
The second term  in Eq. (\ref{fin}) gives for the case of $\Omega_{\mathrm{m}}=0.3$,
\begin{equation}
0.33 \cdot k_E \frac{T_{\mathrm{E}}}{\tau_0} \, = \, 0.33 \cdot 1.60 \cdot 10^5 cm \cdot s^{-1 } \; \frac{86400 \; s}{13.8 \cdot 10^9 \; yr}
 = 0.33 \; \left[\frac{cm}{yr}\right] \, .
 \end{equation}
The direct expansion effect $\frac{R}{t_0}$ is
\begin{equation}
0.33 \cdot  \frac{R}{\tau_0} \, = \,0.33 \cdot  \frac{3.844 \cdot 10^{10} \, cm}{13.8 \cdot 10^9 \; yr} = 0.92  \; \left[\frac{cm}{yr}\right] \, .
\end{equation}
This term corresponds to a third of the general Hubble-Lemaitre expansion.
Summing the various contributions, we get
\begin{eqnarray}
\frac{dR}{d\tau}  = (2.85-0.33+0.92) \,\mathrm{ cm/yr}=
 3.44 \,\mathrm{cm/yr} , \,  \mathrm{from\; historical\; data}\, \cite{Stephenson16}.\label{final} \, .
\end{eqnarray}
Thus, we see that the scale invariant analysis is giving a relatively good agreement with the  lunar recession
 of 3.83 cm/yr obtained  from LLR observations. The difference amounts only to  10 \% of the observed lunar recession. 
   This is clearly much better than in the standard 
 case,  where the predicted  LOD  of 2.395 ms/cy corresponding to the observed recession diverges from the observed one
 of 1.78 ms/cy by 35 \%.  Thus, the scale invariant analysis appears to give more consistent results that the standard case.
We do not consider this as a proof of the scale invariant theory, but  this shows that the SIV theory deserves some attention,
especially since more than several other astrophysical tests (see introduction) are successful. 
 
 We point out that the discrepancy  between the observed LOD from historical records and  the observed lunar recession has
 already been mentioned by several authors.
  Munk \cite{Munk02} has emphasized  the interest 
 of the discrepancy between the LLR lunar recession
 and the LOD data, invoking climatic effects. However, this would not be consistent
  with the data by Deines and Williams \cite{Deines16} which
 cover a much longer period of time  over the geological epochs.
  Dumin \cite{Dumin03} has also clearly demonstrated the above discrepancy between the LLR and the LOD results, 
  showing that  there was an excess of lunar recession  of about
  1.3 cm/yr not accounted by the slowing down of the Earth, 
  an  excess which would correspond to a fraction of about the half of the Hubble expansion.
  Dumin  further  studied this discrepancy  and discussed some possible origins of  it \cite{Dumin16}. Krizek and Somer
  \cite{Krizek16} have supported a local expansion of at least the order of the half of Hubble rate from an analysis of various 
  properties in the Solar System. Also, the reality of the discrepancy between the observed LOD and the lunar recession
  was emphasized by Stephenson et al. \cite{Stephenson20}.

 On one side, we may of course  wonder whether this small effect of about 1.3 cm/yr
 is   sufficient to claim for an additional symmetry property in Physics.  On the other side, steps forward
 often come by scrutinizing minute differences. 
Also, a larger effect would have already been found by more people than the few above precursors.

\section{Conclusions} \label{concl}

We have studied several mechanical properties in the scale invariant context, in particular we have shown the large reduction 
of the additional dynamical effects with the matter density. We have considered
 the two-body problem and the tidal interaction in the Earth-Moon system. 
As examples in the two-body system, independently of tidal effects, there would be a cosmological increase of  0.92 cm/yr for the Earth-Moon distance.

The observed lunar recession  from LLR data amounts to 3.83 ($\pm 0.009$) cm/yr and  the  corresponding  theoretical tidal change of the LOD 
 is 2.395 ms/cy  \cite{Williams16a,Williams16b}. Now, the observed change of the LOD
  since the Babylonian Antiquity is only 1.78 ms/cy,  leading to a significant difference of 35 \%  \cite{Stephenson20}.
  Moreover, the value of 1.78 ms/cy is supported by the data from paleontology over   hundreds million years.
  Such a change of the LOD would  correspond to a lunar recession  of 2.85 cm/yr, instead of 3.83 cm/yr as observed.
  The difference in the lunar recession is well accounted for within the dynamics of the  SIV theory (\ref{final}).  
  \emph{A minima},  the above results shows that the problem of scale invariance is worth of some attention.

\section{Acknowledgments:}
A.M. expresses his gratitude to his wife for her patience and support. 
V.G. is extremely grateful to his wife and daughters for their understanding and family support  during the various stages of the research presented. 
This research did not receive any specific grant from funding agencies in the public, commercial, or not-for-profit sectors.


\section{APPENDIX: A note on the  result of Banik and Kroupa}

A criticism of the scale invariant theory was expressed  by Banik and Kroupa \cite{BK20} with the argument that
``the predicted expansion of the Earth–Moon orbit is incompatible with lunar laser ranging data at $>200 \sigma$''.
They used the following expression for the change of the lunar angular velocity,
\begin{equation}
\frac{\dot{\Omega}}{\Omega}= - \frac{\dot{r}_{\mathrm{SIV}}+(3/2) \dot{r}_{\mathrm{Tide}}}{r} \, .
\label{BK}
\end{equation}
The correct expression in the scale invariant theory (SIV) is according to
the unmodified  expression of the orbital velocity $v^2=GM_{\mathrm{E}}/r$, 
and thus via $ \Omega^2=  G M_{\mathrm{E}}r^{-3}$ one has:

\begin{equation}
\frac{\dot{\Omega}}{\Omega}=\frac{1}{2} \frac{\dot{M}_{\mathrm{E}}}{M_{\mathrm{E}}} - \frac{3}{2} \frac{\dot{r}}{r}\,.
\label{rsiv}
\end{equation}
Now, $\dot{r}$ is given by Eq.(\ref{fin}), where the first term on the right of this equation 
 corresponds to $\dot{r}_{\mathrm{Tide}}$ in (\ref{BK}),
\begin{equation}
\frac{\dot{\Omega}}{\Omega}=\frac{1}{2} \frac{\dot{M}_{\mathrm{E}}}{M_{\mathrm{E}}} -
\frac{3}{2}\, \frac{\dot{r}_{\mathrm{Tide}}}{r} - \frac{3}{2\, r} \, k _{\mathrm{E}} \, \psi \,\frac{T_{\mathrm{E}}}{\tau_0} 
- \frac{3}{2}  \psi \,\frac{1}{\tau_0}\, .
\label{fin2}
\end{equation}
The first and last terms on the right simplify, we get
\begin{equation}
\frac{\dot{\Omega}}{\Omega}= -   \psi \,\frac{1}{\tau_0} -\frac{3}{2} \frac{\dot{r}_{\mathrm{Tide}}}{r} - \frac{3}{2\, r}  k _{\mathrm{E}} \, \psi \,\frac{T_{\mathrm{E}}}{\tau_0} \, .
\label{fin3}
\end{equation}
The first two terms are identical with those  of Banik and Kroupa, however the last term, which is an important one, is absent
in their equation. This clearly invalidate their claim. On the contrary,  the results of the present work confirm the remarkable 
compatibility of the scale invariant theory with the observed lunar recession, an agreement which does not exist in
the standard theory.


\begin{thebibliography}{1}

\bibitem{Azarevich17}
Azarevich, M.B., Lopez, V.L.:
Lunar recession encoded in tidal rhythmites: a selective overview with examples from Argentina
Geo-Marine Letters. \emph{37} (2017),333--344

\bibitem{Baenas21}
 Baenas, T., Escapa, A., Ferrandiz, J.M.:
 Secular changes in length of day: Effect of the mass redistribution
 Astron. \& Astrophys.\emph{648} (2021), 89--98
 
 \bibitem{BK20}
 Banik, I., Kroupa, P.:
 Scale invariant dynamics in the Solar system.
 MNRAS, \emph{407} (2020), L62-66

\bibitem{Bondi90}
 Bondi, H.:
The cosmological scene 1945-1952.
\emph{Modern Cosmology
in Retrospect} (1990) Bertotti, B., Balbinot, R., \& Bergia, S. Cambridge Univ. Press., 189

\bibitem{Bonnor00}  Bonnor, W.B. 
Local Dynamics and the Expansion of the Universe.
Gen. Rel. Grav. \emph{32} (2000) 1005.

\bibitem{BouvierM78}
Bouvier, P., Maeder, A.:
Consistency of Weyl's Geometry as a Framework for Gravitation.
Astroph. Space Sci. \emph{54} (1978), 497-508



\bibitem{Canuto77}
Canuto, V., Adams, P.~J., Hsieh, S.-H., \& Tsiang, E.:
Scale-covariant theory of gravitation and astrophysical applications.
Phys. Rev.D. \emph{16} (1977), 1643-1663 


\bibitem{Christodoulidis88}
Christodoulidis, D. C., Smith, D. E., Williamson,  R. G. et al.:
Observed Tidal Braking in the Earth/moon/sun System.
J. Geophys. Res. \emph{93(B6)} (1988),  6216--6236

\bibitem{Carroll92}
Carroll, S.~M., Press, W.~H., \& Turner, E.~L.: 
The Cosmological Constant.
Ann. Eev. Astron. Astrphys. \emph{30}, 499--542

\bibitem{Deines16}
Deines, S.D., Williams, C.A.:
Earth's rotational deceleration: determination of tidal friction
independent of timescales.
Astron.Journal. \emph{151} (2016). 103--114


\bibitem{Dirac73}
Dirac, P.~A.~M.:
Long range forces and broken symmetries,
 Proceedings of the Royal Society of London Series A,
\emph{333} (1973), 403--418


\bibitem{Dirac74}
Dirac, P.~A.~M.:
Cosmological Models and the Large Numbers Hypothesis.
 Proceedings of the Royal Society of London Series A, \emph{338} (1974), 439--446
 

\bibitem{Dumberry06}
 Dumberry, M, Bloxham, J.:
 Azimuthal flows in the Earth's core and changes in length of day at millennial timescales.
 Geophys. Journal. \emph{165} (2006), 32--46

 
\bibitem{Dumin03} Dumin, Yu.V.:
A new application of the lunar laser retroreflectors: Searching for the ``local'' hubble expansion.
 Adv. Space Res. \emph{31} (2003), 2461-2466

\bibitem{Dumin16} Dumin, Yu.V.:
Local Hubble expansion: current state of the problem.
In \emph{Cosmology on Small Scales} (2016), Eds. M. Krizek and Y. Dumin, Institute of Math. CAS, Prague, 23--40

\bibitem{Einstein45} Einstein, A., Straus, E.G.: 
The Influence of the Expansion of Space on the Gravitation Fields Surrounding the Individual Stars.
Rev. Mod. Phys.  \emph{17} (1945), 120.

\bibitem{Feynman63}
{Feynman}, R.~P.:
Mainly mechanics, radiation, and heat.
Feynman lectures on physics, \emph{1}(1963)


\bibitem{Frie08}
Frieman, J.~A., Turner, M.~S., \& Huterer, D.: 
Dark energy and the accelerating universe.
Ann. Rev. Astron. Astrophys. \emph{46} (2008), 385--437

\bibitem{Gross09} 
Gross, R.: 
Earth Rotation Variations - Long period, 
in \emph{Treatise of Geophysics, vol 3, Geodesy} (2009) Ed. G. Schubert, Elsevier, Amsterdam, p. 239

\bibitem{Gu+M'22}Gueorguiev, V.G.; Maeder, A. 
The Scale Invariant Vacuum Paradigm: Main Results and Current Progress. 
Universe \emph{8} (2022) 213. 


\bibitem{Jesus18} Jesus, J.F. :
Exact solution for flat scale-invariant cosmology.
Rev. Mex. Astron. Astrophys.  \emph{55} (2018), 17--20

\bibitem{Krizek16}
Krizek, M.,  Somer, L.:
Anthropic principleand the local Hubble expansion.
In \emph{Cosmology on Small Scales} (2016), Eds. M. Krizek and Y. Dumin, Institute of Math. CAS, Prague, 65--94



\bibitem{Maeder17a} Maeder, A.:   
An Alternative to the LambdaCDM Model: The Case of Scale Invariance.
Astrophys. J.  \emph{834} (2017a), 194--210 


\bibitem{Maeder17c} Maeder, A.:
Dynamical Effects of the Scale Invariance of the Empty Space: The Fall of Dark Matter?
Astrophys. J. \emph{849} (2017c),158--177


\bibitem{Maeder19}
Maeder, A.: 
Evolution of the early Universe in the scale-invariant theory.
arXiv:1902.10115.

\bibitem{MBouvier79} Maeder, A., Bouvier, P.:
Scale invariance, Metrical Connection and the Motions of Astronomical Bodies.
 Astron. Astrophys., \emph{73} (1979), 82-89

\bibitem{MaedGueor19} 
Maeder, A.; Gueorguiev, V.G.:
The growth of the density fluctuations in the scale-invariant vacuum theory
Physics of the Dark Universe, \emph{25} (2019), 100315--100330


\bibitem{MaedGueor20a}
Maeder, A.; Gueorguiev, V.G.: 
The Scale-Invariant Vacuum (SIV) Theory: A Possible Origin of Dark Matter and Dark Energy.
Universe, \emph{6} (2020a), 46

\bibitem{MaedGueor20b}
Maeder, A.; Gueorguiev, V.G. 
Scale-invariant dynamics of galaxies, MOND, dark matter, and the dwarf spheroidals.
MNRAS, \emph{492}(2020b), 2698

\bibitem{MaedGueor21a}  
Maeder, A.; Gueorguiev, V.G.:
Scale invariance, horizons, and inflation.
MNRAS \emph{504} (2021a), 4005-4014


\bibitem{MaedGueor21b} 
Maeder, A.; Gueorguiev, V.G.:
On the relation of the lunar recession and the length-of-the-day.
  Astroph. Space  Sci. \emph{366} (2021b),  101--125
  
  
\bibitem{McCarthy86} 
McCarthy, D.D., Babcock, A.:
The length of day since 1656.
Physics of the Earth and Planetary Interiors. \emph{44} (1986), 281--292

\bibitem{McVittie32} 
McVittie, G.C.
Condensations in an Expanding Universe.
MNRAS \emph{92} (1932) 500.

\bibitem{McVittie33} 
McVittie, C.G. 
The Mass-Particle in an Expanding Universe.
MNRAS \emph{93} (1933) 325.
 
 \bibitem{Milgrom14}
 Milgrom, M.:
 MOND laws of galactic dynamics.
 MNRAS \emph{437} (2014), 2531--2541
  
\bibitem{Mitrovica11}
 Mitrovica, J.X., Wahr, J.:
 Ice Age Earth Rotation.
 Ann. Rev. Earth \& Planetary Sci. \emph{39} {2011), 577--616


\bibitem{Mitrovica15}
 Mitrovica, J.X., Hay, C.C., Morrow, E. et al.:
Reconciling past changes in Earth’s rotation with 20th century global sea-level rise:
Resolving Munk’s enigma
Sci. Adv. \emph{1} (2015), e1500679-1500684
 
 \bibitem{Munk02}
 Munk,W.H.:
 Twentieth century sea level: An enigma.
 Proc.Natl. Acad. Sci. USA. \emph{99} (2002), 6550--6555.

\bibitem{Sidorenkov97} 
Sidorenkov, N.S.: 
The effect of the El Nino Southern oscillation on the excitation of the Chandler motion of the Earth's pole.
Astron. Reports (in Russian). \emph{41} (1997), 705--708
   
\bibitem{Stephenson16}
Stephenson, F.R., Morrison, L.V., Hohenkerk, C.Y.:
Measurement of the Earth's rotation: 720 BC to AD 2015.
Proc. R. Soc. \emph{A472} (2016), 404--430.

\bibitem{Stephenson20} 
Stephenson, F.R., Morrison, L.V., Hohenkerk, C.Y.:
Eclipses and the Earth's Rotation
 General Assembly,Proceedings of the IAU \emph{IAU XXX } (2020), 160--162

  
\bibitem{Williams94}
 Williams, J.G.:
 Contribution to the Earth's Obliquity Rate, Precession, and Nutation
  Astron. J. \emph{108} (1994), 71


\bibitem{Williams16a}
 Williams, J.G., Boggs, D.H.,  Ratcliff, J.T.:
 Lunar Tidal Recession.
 47th Lunar and Planetary Science Conference (2016a),   1096-1100

\bibitem{Williams16b}
Williams, J.G., Boggs, D.H.: 
Secular tidal changes in lunar orbit and Earth rotation.
 Celestial Mechanics and Dynamical Astron. \emph{126} (2016b), 89--129

}



\end{thebibliography}
\end{document}